# A SMALL, RAPID OPTICAL-IR RESPONSE GAMMA-RAY BURST SPACE OBSERVATORY CONCEPT (THE NGRG)

B. Grossan[1,2], P. Kumar[3], D. Perley[4], G.F. Smoot[1,5]

[1] Extreme Universe Laboratory, Moscow State University, Russian Federation

[2] University of California at Berkeley Space Sciences Laboratory, USA

[3] University of Texas at Austin, USA

[4] California Institute of Technology, USA

[5] University of California at Berkeley, USA

Bruce Grossan: Lawrence Berkeley Lab, 50R-5005, 1 Cyclotron Road Berkeley, CA 94720; Bruce_Grossan@lbl.gov

Pawan Kumar: University of Texas at Austin Department of Astronomy, 1 University Station C1400, 2511 Speedway - RLM 17.204; pk@surya.as.utexas.edu

Daniel Perley: Daniel Perley, 1200 E. California Blvd., California Institute of Technology, 91125, dperley@astro.caltech.edu

George F. Smoot, Lawrence Berkeley Lab, 50-5005, 1 Cyclotron Road Berkeley, CA 94720; GFSmoot@lbl.gov

Abstract

After *Swift*, there is no sure plan to furnish a replacement for the rapidly disseminated, high-precision GRB positions it provides, nor a new type of observatory to probe new GRB parameter space. We propose a new GRB mission concept, the Next Generation Rapid Optical–NIR (near infra-red) Response GRB Observatory (NGRG) concept, and demonstrate, through analysis of *Swift* BAT data, studies of new GRB samples, and extinction predictions, that a relatively modest size observatory will produce valuable new measurements and good GRB detection rates. As with *Swift*, GRBs are initially located with a coded-mask X-ray camera. However, the NGRG has two distinguishing features: First, a beam-steering system to begin optical observations within ~ 1 s after location; second, in addition to the optical camera, a separate near-IR (NIR) camera viewing the same field, greatly increasing sensitivity to extinguished bursts. These features yield the unique capability of exploring the rise phase of GRB optical-NIR emission. Thus far, among GRBs with optical afterglow detections, a peak is measured in only ~26-40% of the light curves. The rise time for prompt, or pre-afterglow optical emission is rarely measured, as is the transition to afterglow emission. Prompt or pre-afterglow NIR emission is even less frequently measured. Rapid-response measurements give new tools for exploration of many science topics, including optical emission mechanisms (synchrotron vs. SSC, photospheric emission) and jet characteristics (reverse vs. forward shock emission, baryon-dominated vs. magnetic dominated). The rapid-response capability also allows measurement of dynamic evolution of extinction due to vaporization of progenitor system dust. This dynamic dust measurement is the only tool we know of to separate the effects of star-system-scale dust and galactic-structure-scale dust; it is remarkable that this probe of small-scale phenomena can be used at the high redshifts where GRBs are observed. In this paper, we discuss techniques and feasibility of these measurements, and give detection rate estimates using only measured *Swift* performance (without extrapolations). The NGRG will explore two new frontiers: optical and NIR GRB emission measured earlier than ever before, via rapid-response, and potentially fainter, more extinguished GRBs than ever before, via sensitive, early NIR measurements.

In an era with little funding for new extragalactic science space missions, costs are important. Our modest NGRG concept will produce new GRB science, while providing crucial access to rapid GRB alerts for the community. An X-ray instrument barely 1/5 the detecting area of *Swift* BAT, 1024 $cm^2$, will yield a significant fraction of BAT's GRB detection rate: more than 65 X-ray detections per year. With a 30 cm optical-IR telescope and modern cameras, more than 19 NIR and 14 optical band detections would be produced each year for community follow-up. In addition, active control of the beam-steering system, via feedback from a fast-read optical camera, would remove the need for arc sec pointing stabilization of the spacecraft platform, for a substantial cost saving and a wider range of potential space platforms.

# 1. Introduction – A Scientific Opportunity

## 1.1. The limits of rapid optical/IR response to gamma-ray bursts

The *Swift* gamma-ray burst (GRB) observatory (Gehrels, et al., 2004) initially detects and determines a rough ($\sigma \sim$ 2-4 arcmin) location of GRBs with the Burst Alert Telescope (BAT) coded-mask X-ray camera (15-150 keV). This rough location is sent to the ground and distributed for follow-up via the Gamma-Ray Coordinates Network (GCN; Barthelmy, et al., 1998) in just a few seconds after trigger; it is also used to point *Swift's* two "narrow field" instruments, the UV-Optical Telescope (UVOT; 17' FOV) and the X-ray Telescope (XRT; 24' FOV), for sensitive follow-up. The narrow-field instruments can then precisely localize bursts (~ arcsec precision), though these positions come significantly later (~ 10 min or more). These precision locations are a special strength of *Swift*, enabling follow-up by almost any type of astronomical instrument, resulting in a wide range of productive scientific activity. Early optical detections, that is, those within the first ~100 s after the GRB trigger, however, are dominated by the UVOT instrument responding to BAT detections. Error! Reference source not found. shows a histogram of response times for UVOT detections with a minimum at around 60 s. By pre-swift standards this is extremely fast, but many GRB have a rise time faster than 60 s. Studies of the best-sampled optical light curves report detection of a peak in less than 40% (Panaitescu & Vestrand, 2008) or 26% (Liang, et al., 2012) of bursts. Our limited knowledge of the rise phase of optical-IR GRB light curves is therefore in large part due to UVOT's finite response time.

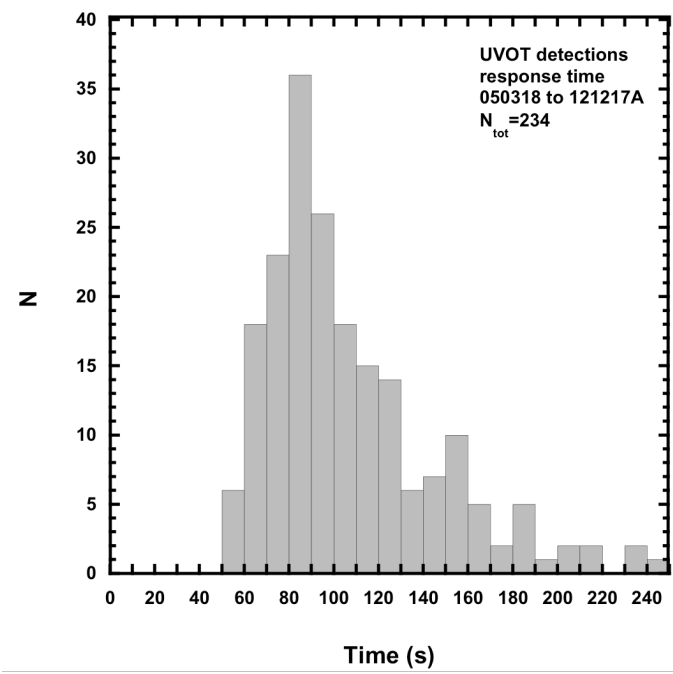

Figure 1 *Swift*-UVOT response time (elapsed time between BAT trigger and start of UVOT observations). Only observations with UVOT detections 2005 Mar. 18 through end of 2012 are shown.

Ground-based instruments (e.g. Super-LOTIS, MASTER-NET, PAIRITEL) have made contributions to rapid response observations, but are also limited in sensitivity and detection rate. The ROTSE program, one of the longest-running early follow-up experiments, has the most published detections with under 50 s response time, arriving on target as fast as ~20 s after a *Swift* trigger (median 45 s; Rykoff, et al., 2009). Unfortunately, the low rate of ROTSE detections and rise-time measurements underscores the difficulties of ground-based rapid follow-up observing. ROTSE announces ~6 upper limits/yr, and ~3 detections/yr (GCN notices 2011 - 2012), compared to *Swift*'s 94 GRB detections/year (Table 1). ROTSE's numbers are low because first, it is very difficult to make telescopes that are both sensitive (i.e. large-aperture) and have short response times. ROTSE-III telescopes respond to alerts in ~ seconds, but have only 45 cm apertures (Akerlof, et al., 2003) and large (3.3") pixels, and are not sensitive enough to detect many bursts. (Larger, more typical and sensitive optical telescopes require several minutes for large moves across the sky.) Second, clouds, daylight, and different accessible sky from the GRB location instrument take a very heavy toll on duty cycle. Co-location with the X-ray location instrument greatly increases the productivity of follow-up instruments. Finally, even at new moon, the background on the ground greatly reduces the sensitivity of optical, and especially IR instruments compared to those in space. The 30 cm UVOT with a ~ 20% QE cathode detector has a sensitivity of ~18.8 mag in 10 s unfiltered (W) exposures (Table 1); reported ROTSE-III sensitivities are ~ 16.9 in R band in 10 s (Figure 2). Even with all its impressive measurements to date, the ROTSE-III experiment has an insufficient sensitivity and response time combination to fully characterize GRB rise times: In Rykoff et al. (2009) 8 detected bursts are presented with response times of 20-60 s; of these, only 2 measurements of a rise time were made. The fastest responses of ROTSE-III instruments, for bursts bright enough to detect, are still not fast enough to characterize the rise phase of most bursts.

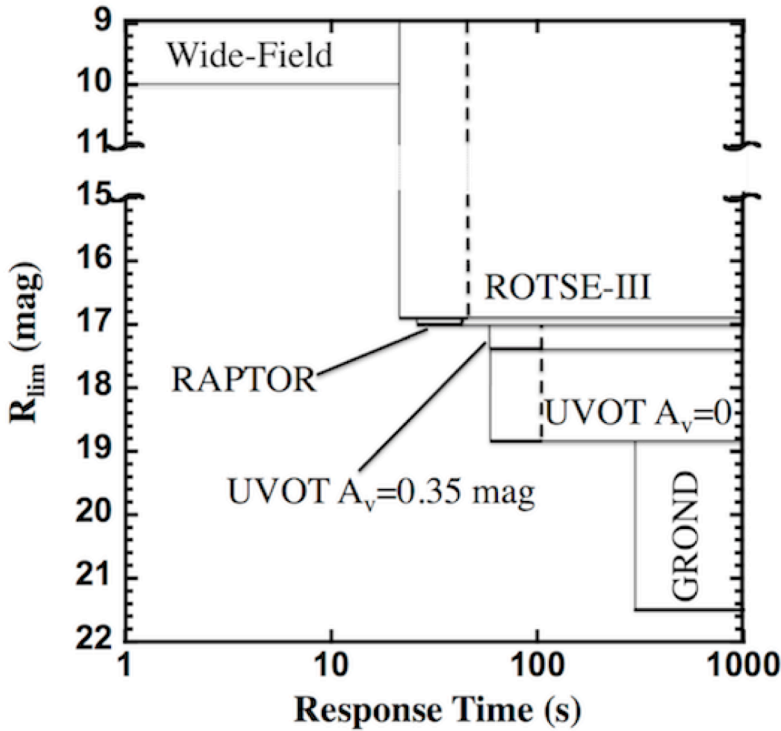

Figure 2 Instrument sensitivity vs. response time. Except for UVOT, values taken from publications or representative samples of GCN reports. Below 10 s there is little coverage at reasonable sensitivity; only moderate sensitivity is available until several hundred seconds, when ~ 2 m class telescopes can respond (GROND). Each rectangle represents the range of response time and equivalent 5 σ sensitivity in 10 s exposure in R, (no conversions were made between, e.g., Rc, R' and Johnson R; sensitivity was scaled by $t^{1/2}$ when 10 s exposures not found). Dashed lines give median response time, left vertical border of rectangle, minimum time. "RAPTOR" refers to the narrow-field version of RAPTOR, and has similar performance to ROTSE-III. UVOT has little sensitivity in R-band; in its standard W filter, most sensitivity is blue of 0.5 μm. Extinction therefore strongly affects UVOT's ability to detect a GRB with a given R flux, as extinction reduces the flux in UVOT's band much more than for that of other instruments. We therefore show "UVOT $A_v$=0", the sensitivity for an unextinguished GRB, R < 18.8 mag (log slope –0.75, galactic $A_v$=0.08 mag, zero source extinction). We also show the sensitivity, R<17.4 mag, for a typically extinguished GRB, "UVOT $A_v$=0.35 mag", with $A_v$=0.35 mag at the source at a typical *z* of 1.8.

Other very-wide field ground-based follow-up instruments can attempt to measure more poorly localized GRB, but have necessarily poor sensitivity due to the large FOV and therefore large sky noise per pixel. The rates of GRB in the required extremely bright regime are very low. The TORTORA instrument made a truly spectacular measurement of the "Naked Eye" burst 080319B (Beskin, et al., 2010), teaching us much of what we know about prompt X-ray and optical emission correlation with ~ 1s time resolution. The RAPTOR full-sky monitoring telescopes detected the early optical emission of 130427A (Vestrand, et al., 2014) with ~ 10 s time resolution, and Pi-of-the-Sky detected 080319B. However, the same instruments have made no other similar quality detections due to lack of sensitivity (11.3-12.4 mag/12 s and ~ 10 mag for TORTORA and RAPTOR full-sky respectively); bursts as optically bright as these are a once or twice-per-decade phenomenon.

We are missing important transient and GRB science without a sensitive, space-based, faster-than-*Swift*-response capability. Generally speaking, the low fraction of optical–IR peak detections reflects a poor understanding of a population; yet these poor detection numbers show that there is a population of optical–IR fast-rising bursts, and we are missing measurements of their rise phase. A new method to measure the rise optical-IR behavior is then a great opportunity, as this information is of great scientific interest (see below, section 1.3).

1.2. The Next Generation, Rapid Optical-IR Response Space Observatory Concept

In order to make a systematic study of the first 60 s of GRB optical-IR (OIR) emission, we explore the concept of a rapid optical-IR response space observatory. Below, we demonstrate that such an observatory can be relatively small, commensurate with currently available resources, and still produce excellent new science. A coded mask aperture X-ray camera is used to detect and locate the GRB (as *Swift* has, but likely smaller). Rapid on-board reduction of the X-ray images, as done on *Swift*, would allow the OIR measurements to begin as early as possible.

In order to make very early OIR observations possible, instead of rotating an entire spacecraft to point a telescope, as *Swift* does, we propose a beam-steering system to provide a much faster response to the X-ray trigger. An OIR telescope beam can be steered simply via the pitch-yaw motion of a flat mirror placed in front of the telescope (**Figure 3**). This beam-steered optical-IR telescope (OITel) is to be instrumented with separate optical and near-IR sensitive cameras that operate simultaneously and view the same field; this is accomplished by means of a dichroic to send separate beams to the different cameras. The X-ray camera and the beam-steering OITel make up the two main instrument systems of this Next Generation Rapid Response GRB observatory (NGRG) concept.

A detailed discussion of the instrument design for the NGRG is beyond the scope of this paper. However, the basics of the beam-steering system have already been demonstrated: a fast beam-steering telescope system has been built, and lab results suggest ~ 1 s slew + settle time for $\geq 35°$ deflections of the beam for a 10 cm aperture / 15 cm beam-steering mirror system (Jeong, et al., 2013). Application of precision motors, light mirrors, and design to minimize settling time can be used to make a much larger system that still achieves pointing on ~ few s time scales. (As mirrors become large, torques increase rapidly; counter-rotating systems might be required so as to not de-stabilize the pointing of the spacecraft platform, but we assume this is a straightforward, solvable, mechanical engineering problem.)

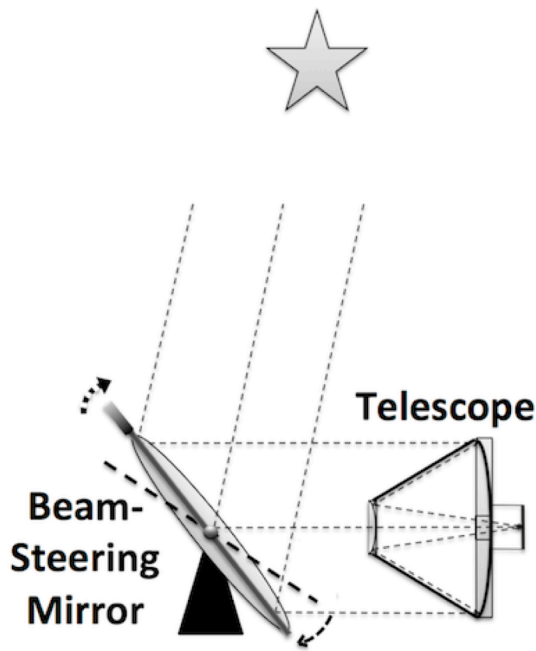

**Figure 3** Beam Steering Mirror Concept. A mirror with two axes of tilt placed in front of a fixed telescope can re-direct the telescope beam within a large range of solid angle. For a light, low moment of inertia mirror, this can be accomplished more quickly than moving an entire telescope or spacecraft, as is done by *Swift*.

The main objective of the NGRG is to systematically measure the first 60 s of OIR GRB emission, and so, to minimize the cost of this proposed observatory, we propose to omit a focused X-ray telescope. Emphasizing early OIR emission studies and omitting a focused X-ray telescope would not abandon afterglow studies or follow-up, however. Given an optical or IR detection, the OITel would provide sub-arc sec quality positions, providing the basis of follow-up observations by other instruments in all wavebands. X-ray follow-up could still be done by other narrow-field X-ray instruments, such as by *Swift*, Chandra, potentially SVOM (Godet, et al., 2012), and others.

An observatory with a smaller, less sensitive, but otherwise similar X-ray coded mask camera to that of BAT would not be expected to detect many new *types* of GRB. We are seeking new science with, for the most part, the already well-studied *Swift* GRB population. The primary areas of exploration are first, the new time regime of < 60 s after trigger in the optical; second, for the first time, prompt and sub- 100 s IR emission will be measured. (It should be noted that RATIR makes IR observations occasionally on the ~ 10 minute time scale, but not on the < 60 s time scale; Butler, et al., 2012) These two primary areas open the doors to many new and important GRB science topics.

### 1.3. Rapid-Response Science

The basic mechanisms of GRB OIR emission at early times have not been positively identified. Since this emission is probably related to the GRB jet, detailed understanding of the origin of this emission gives us information on jet structure, composition, and dynamics.

Before proceeding, we clarify our use of the term "early" emission. Optical emission has a distinct early phase that may plateau (e.g. Rykoff, et al., 2009, Beskin, et al., 2010, Vestrand, et al., 2005) and may have rapid variability like the X and γ prompt emision (Racusin, et al., 2008), but is clearly inconsistent with the power-law decays seen in the optical afterglow phase. The term "prompt optical" emission is usually defined to be emission simultaneous with the initial hard X-ray emision (10's of keV range observed by BAT, GBM, and similar instruments), and can imply a direct relation to high-energy processes; we use the term "early" for OIR emission here to mean bright emission that is clearly before the afterglow phase, but which may or may not be related to emision in the X – γ bands. Immediately after this early phase, optical observations can show a power-law decay indicative of afterglow, but can also show a much more rapid decay (as in the case of a $t^{-6.5}$ rapid decay reported for 080319b; Racusin, et al., 2008) in transition to a slow-decay afterglow.

### 1.3.i Emission Mechanisms

The basic emission mechanisms are not unambiguously established for all GRBs. The relation between early OIR and hard X-ray emission is often a distinguishing feature of emission mechanisms, making observations of OIR emission at early times particularly valuable. We describe features of a few basic mechanisms here. First, we consider emissions with correlated OIR and hard X-ray emission. Synchrotron emission in the internal shock scenario (ISS) is expected to come from the electrons accelerated by interactions of successive shocks resulting in a spectrum rising as $F_\nu \propto \nu^{+1/3}$ for $\nu < \nu_{peak}$, with $\nu_{peak}$ in the ~ MeV range (e.g., Ghisellini, Celotti, & Lazzati, 2000). This mechanism would produce hard X-ray ($\nu < \nu_{peak}$) and OIR emission correlated in time, the one extrapolated from the other with a single spectral index. (Consideration of cooling processes modifies this spectrum to $F_\nu \propto \nu^{-1/2}$ for $\nu_{cool} < \nu < \nu_{peak}$ for fast-cooling, from the hard X-ray band down to $\nu_{cool}$ ~ optical, but this index is not consistent with observed hard X-ray band slopes; Ghisellini, Celotti, & Lazzati, 2000; Piran, 1999; Mészáros, 2002). In the synchrotron self-Compton (SSC) process, a small fraction of seed synchrotron photons peaking (in $F_\nu$ vs $\nu$) somewhere between IR and UV are Compton-scattered to make the much weaker X-γ emission (in $F_\nu$ vs $\nu$) with similar spectral shape for the seed and scattered photons (Zou, Piran, & Sari, 2009; Piran, Sari, & Zou, 2009). A defining characteristic of SSC is that for a given detection in hard X-rays, the (unabsorbed) OIR flux density is much brighter than for synchrotron emission. Therefore, if the OIR emission lies far above the extrapolation of $\nu^{+1/3}$ from hard X-rays, with similar OIR and hard X-ray spectral shapes, then an SSC origin is supported. If the OIR emission lies on the $\nu^{+1/3}$ extrapolation, synchrotron (ISS) is indicated. If OIR is fainter than $\nu^{+1/3}$, both ISS and SSC are ruled out unless the synchrotron self-absorption frequency is above the observed OIR bands. Additional information can also be extracted from OIR and hard X-ray observations: In SSC, the electron Lorentz factor can be determined if peak frequencies of both the seed OIR and scattered X- and γ-ray components are observed; their ratio is given by the electron Lorentz factor (within a factor ~ unity). When the absorption frequency is observed, this gives the distance from the central engine where γ-rays are produced, as well as the electron Lorentz factor and magnetic field strength in the jet (Shen & Zhang, 2009). GRB080319B had optical emission ~ $10^4$ times the flux density in X-rays, much brighter than a $\nu^{+1/3}$ extrapolation downward in frequency from the X-ray band, and rough correlation of X and optical for the first 50 s; the optical emission was therefore interpreted as ISS, and the X-ray emission was ascribed to SSC.

The photospheric emission mechanism (Mészáros & Rees, 2000; Pe'er, Mészáros, & Rees, 2006), whereby γ-ray photons come from multiple inverse Compton scatterings within the Thompson photosphere of the jet, is of great interest because of relatively recent Fermi data and fitting (Ryde, 2004; Pe'er, Ryde, Wijers, Meszaros, & Rees, 2007; Ryde, et al., 2011; Veres, Zhang, & Mészáros, 2013) which show that adding a thermal component signficantly improves the fit in some γ-ray spectra. This mechanism would produce very faint OIR emission, as this would be the Rayleigh-Jeans tail of this thermal emission. Given (unextinguished) OIR emission fainter than that expected for SSC, correlated with γ-band emission, and consistent with a Rayleigh-Jeans spectrum ($v^2$ in the OIR), photospheric production of both spectral components would be strongly supported. With the Fermi LAT and GBM instruments, GRB090902B was found to have a broad modified black body component centered at ~ 290 keV, identified as photospheric emission, which dominated the earliest part of the burst (prompt optical observations were not available; Ryde, et al., 2010).

Now consider OIR emission not correlated in time with hard X-ray emission. Separate mechanisms or locations must be invoked for the two components. Hard X-ray GRB emission shows high variability at all time scales, down to ~ ms, and therefore is often ascribed to internal shocks where fast-moving material from the central engine collides with slower material ejected at an earlier time (Rees & Meszaros, 1994). Bright, beamed OIR emission with uncorrelated variability could then come from reverse shock synchrotron emission, and a $t^{-1.5}$ decay would be predicted (Meszaros & Rees, 1993; Sari & Piran, 1997; Sari & Piran, 1999). A good example here is 130427A, with uncorrelated contemporaneous X and optical emission, and optical emission much brighter than the extrapolation of the X-ray spectrum (Vestrand, et al., 2014). Identification of this mechanism also requires a baryon-dominated jet (a reverse shock traveling into a magnetic jet produces weak emission undetectable in OIR; Zhang & Kobayashi, 2005; Narayan, Kumar, & Tchekhovskoy, 2011; Giannios, Mimica, & Aloy, 2008). Alternatively, OIR emission could come from interaction with the ISM, in which case a decay ~ $t^{-1}$ would be observed. GRB061126 actually displayed this uncorrelated afterglow-like decay beginning during BAT emission starting at least as early as ~ 20 s, while the BAT time for 90% of the flux was 26.8s (Perley, et al., 2008).

Finally, if OIR and hard X-ray emission are similarly variable, but uncorrelated, this would suggest either two separate jets with similar mechanisms, or an as yet unknown mechanism. (There are no well-known and well-documented GRB that fit this description.)

### 1.3.ii  An Independent Bulk Lorentz Factor Measurement

Measurement of the bulk Lorentz factor (BLF) in the GRB jet is an important diagnostic of jet conditions. The interaction of the jet and the ISM often produces an optical and X-ray afterglow peak; a simple, nearly model-independent argument applied early by Molinari et al. (2007) shows that the BLF can be measured from the time of this peak (but see also Nava, Sironi, Ghisellini, Celotti, & Ghirlanda, 2012). As pointed out above, a large fraction of optical light curves record only the afterglow decay phase, i.e. the optical response was too slow to catch the peak. Therefore, the available optical BLF distribution is incomplete, i.e. biased toward low BLFs. Measurement of a correlation (or not) with γ-ray measured BLFs (e.g. Ackermann, et al., 2013, Abdo, et al., 2009), would support (or not) a scenario with emission in the two different bands produced in the same jet. A separate optical BLF measurement would allow comparison of the optical and γ-ray BLF for the same GRB, a test rarely, if ever, made.

### 1.3.iii  Dynamic Dust Measurements In Individual, High-z Bursting Star Systems

Very-rapid destruction of circumburst dust by an early optical-UV flash has been proposed (e.g., Waxman and Draine 2000; Perna et al. 2003). If this process occurs, rapid early-time color and brightness evolution would be observed as the radiation "burns" away the dust, changing the observed color from extremely red to blue with the brightening of the optical emission. Direct detection of this process would open new avenues for studying the GRB environments and progenitors; in particular, only the dust local to the GRB would be destroyed and change extinction properties, allowing separation of local and host galaxy dust effects. This process gives perhaps the only tool to study dust in individual star systems independent of host dust, and because of the brightness of GRBs, it could be used to extraordinary red shifts. Most current observations respond too slowly to fully measure this phenomenon, as dust destruction should happen almost completely 60 s after the burst, again requiring rapid-response.

### 1.3.iv   Relation to Non-Photon or "Multi-Messenger" Observations

Gravitational wave (GW) observatories will be dependent on electromagnetic follow-up for identification, red shift, and other source studies. The most likely detectable source for these observatories is Short-Type GRBs (SGRBs). Yet, the optical detection rate of SGRBs is very poor, as is the number of spectra, with rich information about the source and environment. Fast-response could boost the rate of optical detection as well as the rate of spectral measurements by catching the bursts earlier, when they are brighter. Fast near-IR (NIR) reponse also has the potential to increase detection rates, as some bursts suffering extinction in the optical would be significantly less extinguished in the NIR and therefore detectable. Such detection is critical to understanding SGRBs and to the success of GW science.

Observations (upper limit measurements) are now routinely made of non-photon signals predicted for GRBs: high and ultra-high energy cosmic rays (e.g. ARGO-YBJ; Aielli, et al., 2009), and neutrinos (e.g. ICECUBE; Abbasi, et al., 2009). Fast-response optical observations can test, e.g., Lorentz violations, from the time delay between different energy photons, or between photons and non-photon emission. Such observations would revolutionize astronomy and greatly improve our understanding of black holes, neutron stars, cosmology (e.g. Stodolsky, 2000), and strong field gravity.

### 1.4. Goal and scope of this paper

The goal of this paper is to highlight both the capabilities and feasibility of a modest-sized GRB mission with a rapid-response OITel (the NGRG concept). We give predictions of the performance of the NGRG, demonstrating that significant numbers of GRB may be measured in OIR bands earlier than before, for completely new types of measurements. Our "scaled" detection rate estimates are realistic because (1) they are based on analysis of *Swift* and other actual GRB observations, involving only the known populations of bursts already measured by these instruments, and (2) we estimate the performance of instruments explicitly identical to Swift BAT and UVOT, except with scaled-down collecting area: reduced aperture size for our scaled UVOT, and reduced detector area for the scaled BAT. These "scaling" estimates have no assumptions whatsoever about burst or background behavior or instrument performance, except that future bursts and instruments will be the same as in the past. Finally, we offer discussion on increasing the scientific yield of such an observatory by modifying and modernizing *Swift* instruments beyond this scaling .

## 2. X-ray Detection Rates as a Function of Collecting Area

The first, and rather surprising result of our analysis of *Swift* BAT GRB observations is that a significantly smaller instrument than *Swift* can produce a significant number of GRB triggers with similar location quality. Our method is to scale the performance of BAT by detector area, and to assume that a scaled instrument would operate identically to BAT in every way except detector area. The operation of such an X-ray camera depends critically on the background, which is dependent on the orbital parameters of the spacecraft platform, and even on the construction of the spacecraft and instruments around the X-ray camera (through secondary emission). We also emphasize that operational constraints have a strong impact on duty cycle, and therefore detection rate. It must therefore be kept in mind that such results are valid only for instruments on a spacecraft very similar to that of *Swift* with essentially the same orbit and operational parameters.

GRB detection rate is a weaker function of instrument sensitivity than more typical and nearby populations, those made up of steady sources with a uniform distribution in (nearly) Euclidian space. (For typical populations, Log N (> S), the number of detected sources above a given flux S, $\propto S^{-3/2}$.) Instead, histograms of BAT fluence flatten at low fluence, and histograms of peak flux (e.g. Fishman, et al., 1994) are significantly more flat than $-3/2$ log slope long before the limiting peak flux of the instrument. This reflects the well-known result that GRBs are detectable from very great distances. By analogy to steady sources, the number of sources at a given peak luminosity are limited by the finite volume of the source population rather than by the finite sensitivity of the instrument. Unlike such steady sources where detection number is predicted simply by instrument sensitivity and integration time, transient source detection and location depends on the actual light curve of each burst and the triggering algorithm, as described below.

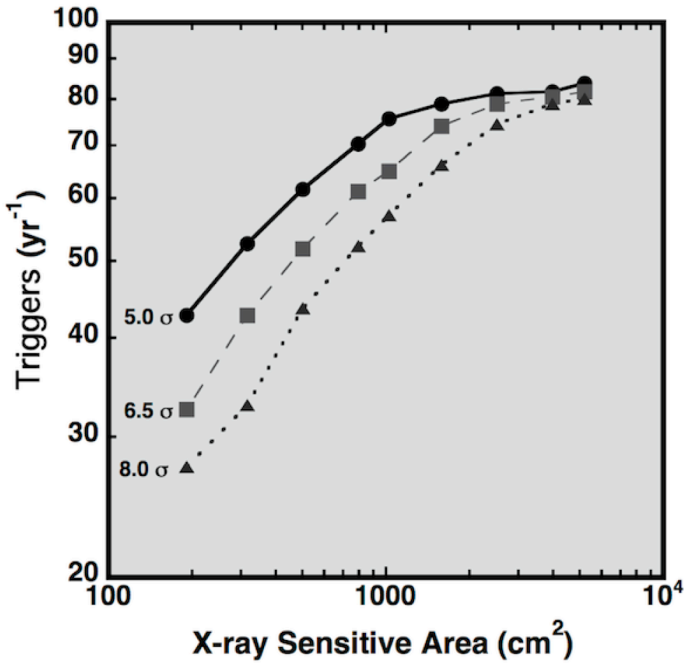

Figure 4 GRB detection rates as a function of BAT collecting area.

2.1. X-ray Triggering Analysis

For simplicity, we use only a rate trigger detection criteria, that is, we determine a source is detected when there is excesss flux above background exceeding a given SNR (signal to noise ratio); we ignore other, less common types of triggers. For a steady source on-axis, SNR ~ S $A^{1/2}/B^{1/2}$, S the source flux, B the background, and A the detector collecting area. Detection depends on only one characteristic of the source, S, and is more weakly dependent on detector area, A, by the 1/2 power. For transient sources, knowledge of the light curve is required; a trigger occurs if any part of the light curve exceeds the SNR criteria in any time bin(s) analyzed. We examined a large sample of BAT light curves for this purpose, to measure the fraction of bursts a "scaled-BAT" of a given size would have been triggered and detected, yielding detection rate as a function of collecting area.

GRB rate as a function of A was determined by (i) finding the peak SNR time segment in each BAT light curve, (ii) scaling the BAT $SNR_{peak}$ by collecting area to get $SNR_{peak}(A)$, then (iii) counting the number of bursts with $SNR_{peak}>$ threshold. We used a very simple $SNR_{peak}$

"trigger", as follows: We used the sum of 64 ms data channels 1-3 (15-100 kev, the highest SNR combination). Integration time windows of 0.25, 0.5, 1, 2, 4, 8 s were examined for fluctuations > threshold (in σ) over background (the trigger/detection criterion). The trailing average background (t–19.2 to t–6.4 s) was used, a reasonable choice for an actual flight trigger. All triggers were checked by eye for false triggers. BAT also has a long time window, or "image trigger", which contributes a relatively small (< 10%) fraction of bursts. We did not use such a trigger, as the benefit for a small instrument would likely be small.

We selected a time period 2006 May 2 – 2008 Oct 7 based on optical observations (explained below), and analyzed 224 BAT light curves in this period. For each light curve, we found the $SNR_{peak}$ in all time windows. We then scaled the SNR for smaller collecting area, and reported the number of bursts with any time bin over trigger threshold with these smaller areas. The result is given in **Error! Reference source not found.**. The number of triggers closely follows $A^{1/2}$, falling off at ~1500 cm$^2$. We also analyzed another 94 GRB light curves 2010 Nov. - 2012 Mar., to check for variation (and found none). Our list of GRBs was taken from the *Swift* GRB Lookup Table (http://swift.gsfc.nasa.gov/archive/grb_table/).

We pre-selected only burst data for this analysis. The main shortcoming of this procedure is concern over false alarm rate: for an actual operating observatory, too many triggers due to noise or background fluctuations would render the instrument useless. In actual operation of the X-ray camera and imaging system, however, there are a number of factors which reduce the false alarm rate. The excess flux in a time window triggers imaging, where the on-board computer deconvolves the detector array signal (using the mask pattern) to form an image of the sky. In the case of a true GRB, a new point source is found on the image; in the case of random fluctuations, a point source is not formed and the trigger is rejected. Known tools are available to control excessive numbers of triggers as well: trigger parameter tuning, cutoff rigidity maps (i.e. when the background is predicted to be high, trigger criteria are increased appropriately), using the knowledge of when bright sources will suddenly enter the FOV, count rate consistency checks between different sections of the detector array, and other techniques. We find no reason that the false alarm rate would strongly increase with decreased detecting area as long as a SNR criteria trigger and other standard tools are used.

### 2.2. X-ray Results

Our simplified trigger detected 93% of 224 BAT bursts in our sample period; 8 of these were image triggers or ground analysis bursts, and 7 other bursts were not detected by our algorithm, mostly due to data gaps interfering with the algorithm. We give results for various values of X-ray detecting area in Table 2. It is remarkable that, for a very small X-ray detector, 190 cm$^2$ of collecting area, more than 27 GRB/yr would still have $SNR_{trig} > 8$.

## 3. Rapid-Response Optical/IR Detection Rate Estimation

The purpose of this section is to make realistic estimates of the rate of OIR GRB detection for a space platform instrument as a function of OIR sensitivity, for response times smaller than that of *Swift* UVOT. In order to make rate and performance predictions for optical detections of GRB, we again look to the extensive data available from *Swift*.

### 3.1. UVOT Early Brightness Distribution

We begin by defining a sample of UVOT observations that describes the earliest possible optical behavior of GRBs.  In the interval 2006 May 2 to 2008 Oct 7, Swift burst response was uniform, with initial ("finding chart") exposures of around 100 s in W, or the unfiltered instrument band (Breeveld, Landsman, Holland, Roming, Kuin, & Page, 2011; Table 3). After 2008, this exposure was changed to around 147 seconds (Table 3), including somewhat later behavior, so we concentrate on the former interval. (Although shorter exposures could be produced by custom reductions, we used only the full exposure results given in the GCN notices.) The slight difference in selections made no significant difference in the distributions of brightness or detection rate. Both the exposure time and the response time after trigger vary; we define our sample with a cut on the time of the mid-point of the exposure after trigger, $t_{mid}$, to be < 170s to retain most bursts, but cut observations with slow response. We refer to the distribution of brightness for W filter/ ~100 s exposure / $t_{mid}$ < 170 s as the "Early UVOT Brightness Distribution" (Figure 5).  This is a reasonable estimate of the distribution of optical brightness of GRBs at ~ 110-170 s. (We note uncorrected non-uniformities in our Early distribution: BAT trigger criteria changed during the life of Swift, UVOT sensitivity changes slowly with time and background, which varies with ecliptic latitude and other factors, but we find these to be acceptable for our estimate and comparison purposes.)  This distribution should be the most sensitive and *systematic* survey of GRB early optical brightness, because equally early ground-based measurements are rare, as they are severely hampered by weather, different accessible sky than any hard X-ray band detection instrument, and other factors. The Early distribution can be seen to have quite low rates below 15 mag, and then to increase ~ factors of 1.5 per mag until a flattening may be seen around 18.4 mag.

The Early Distribution should not be used directly to estimate the rate of GRBs with optical emission. UVOT detects 30.4 GRB yr$^{-1}$ from real-time BAT X-ray detections (Table **1**), compared to the 18.1 yr$^{-1}$ in our Early sample (Table 3). The "missing" GRBs had a slow response by UVOT that would pollute our early observations with late-time behavior. For example, UVOT did not begin observing GRB 130420A until 739 s after the burst (though GCN 14406 states, "*Swift* slewed immediately to the burst"). The Early Distribution explicitly includes operations-related and other effects not relevant to astrophysics that change the observed rates.

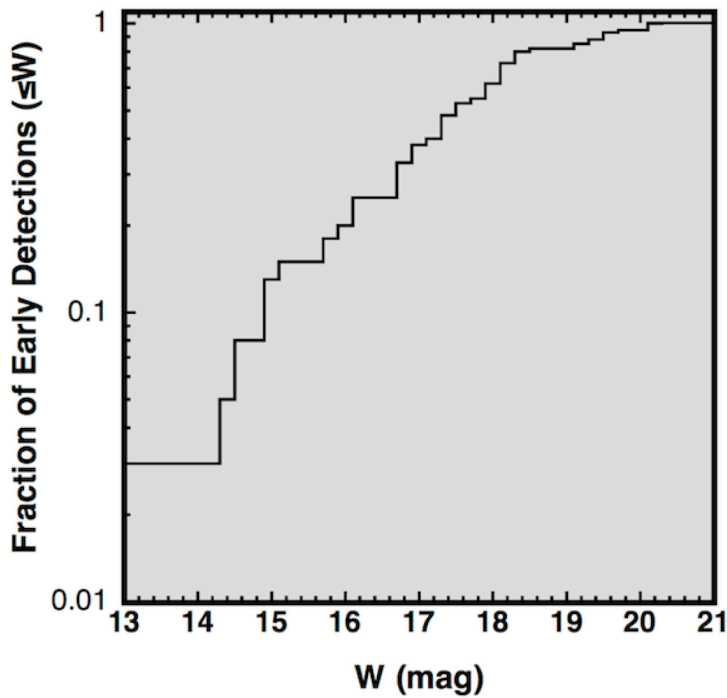

Figure 5 **Early UVOT Brightness Distribution.** See text for explanation and sample definition.

## 3.2. Rapid Detection Criteria

What would it take to improve the measurements or limits of optical rise times of GRBs over that of *Swift*? In this discussion, times are time from BAT trigger, and we ignore early flaring behavior. We make the very simple assumption that one flux followed by a higher one followed by a lower one indicates a peak, and we concentrate on the early-peaking population of bursts that *Swift* BAT+UVOT detected, but failed to observe early enough to measure a peak. Consider an instrument that could measure fluxes at least as faint as UVOT can for $t_{new} < t_{UVOT,earliest}$, the earliest time of UVOT detection, and with an exposure time, $t_{exp} < t$ (i.e. with some kind of useful time resolution). In this case, for the early peaking population, an improved measurement or limit would always result. There are three possible cases: (i) the earliest flux is the same or brighter than the earliest UVOT measurement, in which case you learn that $t_{peak} \leq t_{new}$, a more strict upper limit to the peak time than previously possible; (ii) the earliest flux is fainter than the UVOT detection but detectable by the putative instrument, here a $t_{peak}$ measurement results; (iii) the earliest flux is too faint to measure, and so a $t_{peak}$ measurement (i.e. peak detection) also results. In the latter two cases, a peak is detected, and restricted to the interval $t_{new} < t_{peak} < t_{UVOT,earliest}$. We therefore define a useful measurement as one where the given instrument could detect fluxes fainter than the UVOT earliest flux at a time < 140 s, the <$t_{mid}$> value of the Early Sample. This criterion allows us to estimate conservative useful detection rate *lower limits* based on sensitivity of the optical instrument. These are necessarily lower limits because bursts that were fainter than UVOT's limit at all times might be detected by a more sensitive instrument, and in addition, bursts that peaked above UVOT's limit, but earlier than it could respond, might also be detected in earlier observations, increasing the overall detection rate.

## 3.3. Rate Estimate Results

### 3.3.i Optical Detection Rate Estimates

The Early Brightness Distribution above shows that UVOT can typically detect GRB optical emission to ~20th mag in W in about 100 sec (maximum 20.3 mag). In the sub-100s regime, more time resolution is required. Scaling noise by $t_{exp}^{1/2}$, and assuming a constant spectrum, we find that a UVOT-like instrument would be able to detect GRB to about 18.75 mag (maximum 19.05 mag) at $t_{exp}$ =10 s exposures; such an instrument would be able to place a useful rise time limit on the vast majority (82%) of the Early Sample. (We selected a fiducial 10 s as a minimum time resolution for sub-60 s observations for the rate predictions, below. However, for our proposed system, very fine time resolution is always available, subject only to SNR requirements. UVOT's microchannel plate intensified CCD detector (ICCD), or modern EMCCDs, can be read at sub-second frame rates; they have negligible effective read noise so frames are summed with SNR~ $N_{frames}^{1/2}$, providing very high time resolution for bright bursts with little noise penalty for fainter bursts that require addition of more frames.)

From the Early Sample data, we estimate early optical measurement rates. We use the X-ray SNR ≥ 6.5 criteria, slightly relaxed from the typical SNR ≥8 criteria used by BAT (Barthelmy, et al., 2005). A slightly higher false alarm rate might result, but for this small observatory, with lower event rates, this would be acceptable. There is also a decrease in burst location accuracy for the lowest SNR events, as the source position accuracy is proportional to 1/SNR (Caroli, Stephen, Di Cocco, Natalucci, & Spizzichino, 1987), but < 10% of such bursts would then be off the camera field. In Table 4 and **Figure 6** we summarize the results of these calculations. For an X-ray detector of 1/5 the area of *Swift*, 1024 cm$^2$, at least 11.8 bursts yr$^{-1}$ would be detected optically with a UVOT-like optical instrument; Using 60%-74% non-detection of the rise phase (sec. 1.1), ~ 7-9 improved measurements of rapid-rise GRB would therefore result each year. We take this 1024 $cm^2$ X-ray collecting area, 30 cm optical aperture configuration as our "straw man" small instrument (but still cover other sizes in our calculations).

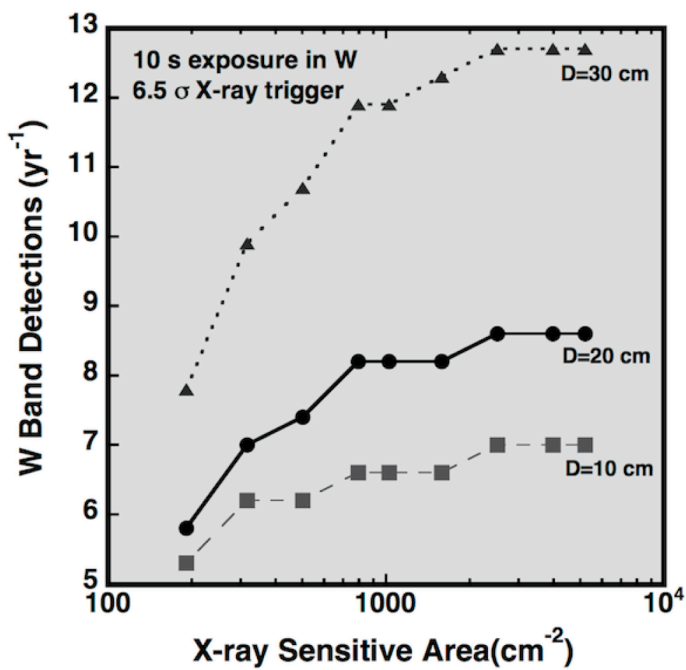

**Figure 6** Rapid Optical Detection Rate. "W" is the UVOT's unfiltered response. Detection rates for different optical aperture diameters, D, are given vs. X-ray instrument detector area. See text for detailed description.

Correlations between late optical and X-ray emission of GRBs have been found (Gehrels, et al., 2008; Panaitescu & Vestrand, 2008), and so one might expect that small instruments with X-ray bright GRBs might have optically brighter bursts, and therefore a higher optical detection rate. In practice, we see only a small effect in detection rates. The early detection fraction for a 5200 $cm^2$ X-ray camera would be 8.5% / 15.5% for a 10 cm / 30 cm optical aperture. For a 1024 $cm^2$ X-ray camera the rate would be 10.1% / 16.5% for a 10 cm / 30 cm optical aperture.

## 4. Extending Swift's Capabilities and Event Rates

### 4.1. Improving Optical Detection Rates with CCD Detectors

The UVOT front surface detector is a bi-alkalai metal cathode, with a very poor quantum efficiency (QE) compared to modern CCD devices, including fast-read devices such as EMCCDs which can operate similarly to UVOT (Table 5). The UVOT cathode QE peaks at ~ 24% at ~3000Å, but rapidly falls below 10% by 5500Å. CCD devices may have QE above 90% from ~ 4100Å-9000Å, yielding dramatically higher sensitivity in a more red band, where GRBs are brighter. Consider, as a likely example, a GRB with an OIR log slope of –0.75 for early emission (Rykoff, et al., 2009), a typical SMC-type reddening of $A_V$ = 0.35 mag (the median of the sample in Covino, et al., 2013), a "typical" high-latitude milky way reddening of $A_V$ = 0.08 mag, $z$=1.8 (median of the sample in Sakamoto, et al., 2011). We estimate that an EMCCD would be about 1.7 mag more sensitive in a 10 s exposure, for the same detector front-surface pixel size, aperture size, etc. Because we know UVOT's sensitivity in longer exposures, we can predict the sensitivity of a high-QE detector in short exposures by scaling, and then predict the effect on rate from the Early Distribution. These rate predictions are given in the "CCD" rows of Table 4. Using a CCD in our "straw man" conuration, ~1/5 the X-ray detector area of *Swift*, with a 30 cm OITel, about 14 GRB/yr. would be detected optically, ~ 8-10 of which would yield improved measurements of fast-rising bursts.

### 4.2. NIR Capability: Increasing Event Rate by Observing Extinguished Bursts

NIR observations from space have revolutionized astronomy due to the lack of atmospheric background. To estimate NIR sensitivity to GRB for this space instrument, we assumed the same GRB spectrum as in the previous section, a camera with a 0.9 – 1.8 μm band, the QE and noise for an H2RG sensor (Table 5) at 155K (Beletic, 2008), the same pixel and aperture sizes as in the previous section, and we assumed zodiacal light dominated the background. The OITel NIR instrument would then detect a typical GRB about 2.8 mag fainter than possible with the UVOT in 10 s, to $W \leq 21.6$.

UVOT has little response red of 0.6 μm, so the difference in detection rate will be dramatic for extinguished bursts, a large fraction of the population. In a sample of 29 *Swift* GRBs (Cenko, et al., 2009), many were found to be "dark", i.e. they had unusually weak optical emission compared to their X-ray flux during the afterglow phase; however, most of these bursts could be detected by either deep or early R, or NIR imaging (Perley, et al., 2009). In order to estimate an increase in detection rate for a NIR camera, we selected bursts from the sample that UVOT responded to but failed to optically detect, but that also had an otherwise measured OIR flux that could be detected by our NIR instrument. We found 7 such bursts (050713A, 050915A, 060210, 060510B, 080320, 070208, and 070419A; all but the last two classified as "dark") with estimated extinctions $A_V \sim 0.5 - 5$ mag, except for 060510B and 080320 which have Lyα absorption in optical bands (Perley, et al., 2009). All but 050915A could be detected in 10 s exposures with any aperture greater than or equal to 10 cm; 050915A, with the limited measurements available (i.e. not the peak brightness) would require a 100 s exposure and a 30 cm aperture for detection. UVOT detected a total of 10 bursts in the sample, indicating that NIR capability would increase the detection rate by at least 17/10 (or by 16/10 for a 10 cm aperture; again without assuming the bursts would become brighter in our earlier observations). Without a clear correlation between BAT brightness and extinction, we therefore apply a simple scaling (by 17/10 or 16/10 as appropriate) to the optical detection numbers in Table 4 (results in Table 6).

### 4.3. Improving X-ray Detection Rates with Updated X-Ray Detectors

Only scaled-down versions of BAT were considered in this work, in order to produce conservative and credible rate estimates. However, implementation of this instrument with more modern X-ray technology would yield significant benefits. *Swift* BAT's nominal range is 15-150 keV, but, Si detectors and newer applications of CdTe and CZT detectors have good sensitivity down to at least 5 keV (e.g. Burrows, et al., 2012, Triou, et al., 2009; Table 5). This would make an X-ray coded mask camera significantly more sensitive: From 5-150 keV, compared to BAT's 15-150 keV, the average long GRB spectrum (Sakamoto, et al., 2011) gives a factor of 5.8 more source photons in the lower-threshold band. The DXRB is flatter than the GRB spectrum in the added 5-15 keV band (e.g. Zombeck, 1990), so unless there was high instrument noise in this added part of the spectrum, greater sensitivity must result. For scaled-down instruments, this would produce higher rates of GRB detection, and in turn, higher rates of OIR measurements.

## 5. Dynamic Measurements of Dust Evolution

In addition to the theoretical predictions of rapid dust destruction in GRBs (Section 1.3.iii), some observations support these predictions. X-ray afterglow observations often show significant gas absorption columns (equivalent $N_H \sim 10^{22}$ cm$^{-2}$; e.g. Galama and Wijers 2001; Stratta et al. 2004; Schady et al. 2007; Perley, et al., 2009), yet typical IR-UV observations show less extinction than predicted from these columns using typical Local Group dust-to-gas ratios (e.g. Prochaska et al. 2009). Evidence for photodestruction of a modest amount of dust ($\Delta A_V \sim$ 0.6 mag) has been presented (Morgan, et al., 2013), providing evidence that this process does occur, and can be observed. The modest change in the extinction reported may be due to responding too late, and catching only the very end of the process. Immediately below, we outline how this process can be more directly and convincingly measured.

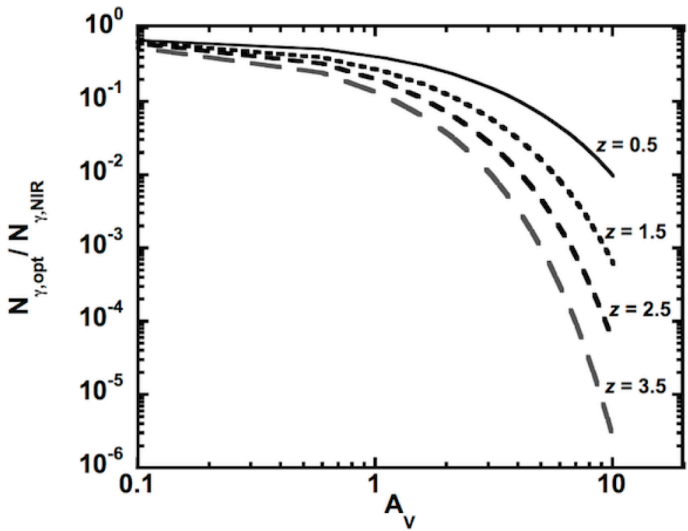

**Figure 7** Optical/IR photon flux ratio vs. extinction. The figure above shows, for various values of $z$, the ratio of optical to NIR band photon fluxes for a given $A_V$. A –0.75 log slope GRB and a $\lambda^{-1.45}$ extinction law were assumed. For $z>2.28$ these relations hold only if no Lyman alpha absorption is present.

Extinction may be estimated via a ratio between bluer, more extinguished bands, and redder, less extinguished bands, so we consider the use of a typical CCD sensor optical camera and an H2RG IR sensor camera to provide a blue and a red band. CCDs are very efficient from at least 4000 Å – 9000 Å; H2RG sensors are sensitive from ~ 7000Å to > 2.5 μm. However, mirrors near 300K (a mirror temperature without special cooling requirements in spaceflight) tend to cause high background in IR cameras for λ > 2.0 μm, so we selected an IR sensor with a 1.8 μm cutoff. We therefore choose a transition to give a very wide band in both detectors for high SNR: an optical band 0.35 μm – 0.9 μm and an IR band (0.9 – 1.8 μm).

We take the early OIR emission of the GRB to be a –0.75 log slope power law, modified by an extinction curve. Extragalactic extinction curves vary in several ways, but seem to follow some general trends: *roughly* speaking, they are similar to a power law of wavelength; only a small fraction of galaxies from studies of active galactic nuclei (AGN; Pitman, Clayton, & Gordon, 2000; Grossan, Remillard, Bradt, Brissenden, Ohashi, & Sakao, 1996) and GRB (Zafar, Watson, Fynbo, Malesani, Jakobsson, & de Ugarte Postigo, 2011) have a 2175 Å feature, a large bump superimposed on this power-law-like curve (e.g. Pitman, Clayton, & Gordon, 2000). To approximate "typical" extinction, we assume an extinction curve similar to the average SMC extinction function (Prevot, Lequeux, Prevot, Maurice, & Rocca-Volmerange, 1984), with extinction $A_\lambda \propto \lambda^{-1.45}$ and no 2175 Å feature. In **Figure 7**, we plot the ratio of the filter bands vs. the extinction, for systems at various red shifts. (We note that for $z>2.28$ Lyman alpha crosses 4000 Å, the blue end of CCD sensitivity, and so the ratio becomes even smaller if neutral hydrogen is present.) Any GRB with an OIR detection would have a high-quality position which would be rapidly broadcast to ground; $z$ could then be determined via absorption features in the burst spectrum, or by observation of the host, preferably by spectroscopy, but also by photometric techniques. Therefore, the ratio of the two bands, plus a red shift, directly gives the extinction at the source for an assumed extinction curve. Monitoring the ratio of these bands during the first ~ minute will then describe the phenomenon of dust vaporization, possibly for the first time. Additional information on the form of the extinction curve could be obtained by using a larger number of filters, or spectroscopy.

# 6. Discussion

### 6.1. What is the Minimal Productive NGRG?

Much of this paper is dedicated to making detection rate predictions as a function of collecting area for the NGRG concept. This reflects our emphasis on detection rate, both overall, and "new types" of detections (at very early times, in IR bands, etc.), as the critical metrics for the success of a future GRB observatory. There are essentially no limits to how large the collecting area of the instruments might be and still increase performance; we wanted to explore the question of how small or inexpensive an instrument could be and still be productive, and provide new capabilities.

We can say that at our "straw man" instrument size (1024 cm$^2$ X-ray detector area and 30 cm OIR aperture; Tables 2, 4, and 5) the NGRG would be very productive, as our realistic and conservative rates (including space operations and background effects, derived from real data) are a significant fraction of those from *Swift*. Considering a nominal mission lifetime of 5 years, about 95 NIR GRB detections are predicted, enough to measure fairly detailed distributions of early OIR emission. The rates are still useful even when considering only studies of the fast-rising population: after 5 years of operation, our straw man configuration would yield 35-45 improved optical rise time measurements (40-50 with CCD detectors); the NIR camera, assuming optical light curve rise times, should yield improved rise time measurements in ~57-70 GRB. These numbers would provide enough bursts to yield some diversity in GRB type, allowing the study (or identification) of some smaller populations. We emphasize again that these rates are conservative; the actual rates could be significantly higher, as OIR detection rates would be increased by earlier measurement, when most sources are brighter. The small X-ray detector area version of this mission concept can be viewed as giving up some fraction of the *Swift* rate, but still retaining the opportunity to study the *Swift* population in new ways, with rapid OIR response and an additional NIR capability to study extinguished bursts in greater detail than before. This is the essential "tradeoff" of the small instrument choice.

We did not find other parameters relating to size or expense that could be significantly changed without reducing the productivity of the mission. In particular, our rates are based on *Swift* performance in a nearly equatorial (20° inclination) low earth orbit (LEO). The X-ray instrument detection rate is sensitive to instrument background rate, and spacecraft in high inclination LEO orbits spend more time in high background regions than those equatorial orbits. X-ray observations cannot be made in high background regions (and for some time after, due to activation of the spacecraft materials), resulting in large reductions in the duty cycle of the observatory, and therefore detection rate. A high-inclination/low duty cycle orbit causes a tradeoff for detection rate, and may not be suitable, even for a substantial cost saving. In the next section, however, we suggest a way to use a less expensive spacecraft than that used for Swift without impacting event rate.

### 6.2. A Simpler Spacecraft Via Imaging-Feedback Mirror Control

Space observatories such as *Swift* and *Hubble* have stabilized pointing so that long exposures are possible with high spatial resolution instruments without significant "trailing" or "smearing" of the image. Such image degradation divides the signal intended for single pixels over multiple pixels, reducing SNR, and causes light from trailed bright sources to overwhelm the signal from faint sources, and other problems. Precision stabilization, however, is extremely expensive and such stabilized spacecraft are rare: a few to a dozen LEO satellites or spacecraft are flown each year, typically stabilized to ~ few arc minutes; because of the expense of ~ arc-second stabilized pointing, only a few such platforms are flown per decade. Fortunately, a variety of methods are available for stabilizing images from a roughly stabilized platform.

In section 4.1 we proposed the use of EMCCDs for optical detectors, which are typically read out at a period << 1 s. For integration times of seconds, and for point-spread function (PSF) centroid calculation times up to a few hundred ms, several cycles of fine-adjusting the beam-steering, analyzing the resulting PSF, and using the result as feedback for the next beam-steering adjustment are possible. This process allows standard feedback control of the beam-steering system, yielding high-quality images without expensive precision spacecraft pointing. We give some technical details and feasibility arguments for this technique in Appendix Section II. Implementing this stabilization would make the NGRG deployable on a wide variety of less expensive space platforms, and therefore, substantially increase its opportunities for flight.

### 6.3. Cooled Mirrors for Improved IR Response

In the mission concept presented here, for cost saving, the telescope mirrors are not cooled, and so we specified a cutoff in the NIR band at 1.8 μm to avoid thermal mirror background. At a typical *Swift* GRB red shift of 1.8, the NIR camera views only wavelengths to the blue of 6430Å, R band, in the burst system frame; if the instrument could observe longer wavelengths, it would have better sensitivity to extinguished bursts at high-$z$. For a slightly cooled mirrors (200k), achievable through passive cooling, the same sensor could be used with great sensitivity out to 2.2 microns; with more aggressive cooling of the mirrors to 100k and the sensor to ~ 75 K, the same sensor could be used to 5 microns. (See Table 5 for cooled mirror instruments.) In the latter case, J and K band in the burst system frame would be observable to $z$~1.8 and $z$~1.1 respectively. (JWST will able to observe these wavelengths, but will not be able to carry out rapid-response observations.)

### 6.4. Conclusion: A Conservative, Yet Productive, Way Forward For GRB Studies

*Swift* is far past its design lifetime, and replacement with any instrument that would allow the continuation of detailed multi-band GRB follow-up by the observing community is far from certain. Fermi and other observatories will continue to make GRB observations, but their instruments do not provide high-quality positions, and the observing community would only rarely be able to make follow-up observations. We argue that community follow-up is critical to progress, because it is actively adaptable to the scientific ideas of the moment, in a way that a space observatory can never be. Given high-quality positions, if a certain type of observation becomes important for a given new line of scientific inquiry, the required ground-based instruments can be brought to bear (within response time limitations); in contrast, space observatories generally *never* change instruments (Hubble is a unique exception). New ground-based instruments can even be built as needed in a year or two, compared to a typical decade or more for a space observatory instrument. We argue that a GRB observatory that provides rapidly-disseminated and precise positions is a unique and critical infrastructure required for a vast range of future GRB work; we point to the productivity of *Swift* follow-up observations as the support of this argument.

Proposed GRB missions to "replace" *Swift* (in that they rapidly disseminate precise positions) include EXIST (Grindlay, et al., 2010) and JANUS (Burrows, et al., 2012). These missions were in large part motivated by the aim to study a new population of GRBs, and were then "optimized for detection of high-z GRB" via emphasis on improved low-energy X-ray response. While it seems obvious that a negative log slope spectrum, high-$z$ source would be more easily detected with good low-energy response, there is no clear correlation of $z$ and any property such as peak energy, duration, etc. in measured data; the broad distribution in intrinsic properties is thought to dominate effects of a wide range in $z$ (e.g. Sakamoto, et al., 2011). These proposed missions are also quite large and costly. The NGRG concept provides a good alternative to larger proposed "*Swift* successor" missions. The NGRG provides high-quality positions, but also breaks new ground, exploring early optical/IR emission of GRBs in a systematic way for the first time, a focus with a clear theoretical and observational foundation. We have shown here that even at modest X-ray instrument size, significant detection rate, and therefore productivity would result. In this way, an NGRG can be presented as a more conservative project both scientifically and fiscally, while still providing the rapid GRB positions that are essential for OIR follow-up science. The Chinese-European mission, SVOM, has 1024 cm$^2$ of X-ray detecting area (the same as our "straw man" size) and an optical camera, but no IR camera, greatly reducing its sensitivity to extinguished GRB. This mission, proposed more than a decade ago, has failed to get a formal approval or start as of this writing. If some eventual successor to *Swift* is not flown, OIR observations will depend on the remaining space observatories for initial detection, all with inferior quality positions, slower reporting, and greatly reduced event rate, leaving few opportunities for studies of early OIR emission and community follow-up.

Acknowledgements:

*This work was supported, in part, by a "MegaGrant" from the Ministry of Education of the Russian Federation, for operation of the Extreme Universe Laboratory at Moscow State University.*

*This research has made use of data obtained through the High Energy Astrophysics Science Archive Research Center Online Service, provided by the NASA/Goddard Space Flight Center.*

The authors wish to thank the following students at Moscow State University, B. Goncharov, G. Rozhkov, K. Saleev, and E. Grobovskoj for their work in obtaining and checking data. We thank Paul Connell for sharing his expertise during numerous discussions and exceptional work on projects leading up to this one. We also thank Nikolay Vedenkin for helpful discussions on spacecraft and instrument communications.

# APPENDIX

## 6.1. Appendix Section I.
**Table A1** *Swift* GRB Measurement Rates

| Start Time | End Time | Parameter | Value | Rate (yr.$^{-1}$) | Note |
|---|---|---|---|---|---|
| 2005 Jan 1.0 | 2013 Jan 1.0 | BAT detections | 736 | 92.0 | a |
| 2005 July 1.0 | 2013 Jan 1.0 | UVOT observations | 600 | 80.0 | b |
| 2005 July 1.0 | 2013 Jan 1.0 | " ", no ground anal. events | 581 | 77.5 | b |
| 2005 July 1.0 | 2013 Jan 1.0 | UVOT detections | 228 | 30.4 | c |
| 2005 July 1.0 | 2013 Jan 1.0 | " ", no ground anal. events | 224 | 29.9 | c |

Source: http://heasarc.nasa.gov/docs/swift/archive/grb_table/

Notes
a. The first publicly released burst was 041217; we chose some short time after that such that operations would be relatively stable. All bursts, including those detected in ground analysis, are included.
b. The first publicly released UVOT observation was of 050124; we chose some short time after that such that operations would be relatively stable.
c. The first publicly released UVOT detection was 050318; we chose some short time after that such that operations would be relatively stable.

## 6.2. Appendix Section II

### 6.2.i    Image Stabilization for an NGRG

Here, we give some technical details and feasibility arguments for the active feedback image stabilization system, using numbers representative of typical low-Earth orbit Earth observation platforms. (Our numbers come from our experience working on the Lomonosov spacecraft, a variation of the Kanopus spacecraft bus.) Nominally the X-ray camera would point radially outward from the center of the orbit, so the optical camera pointing is within 30˙ of the orbital plane. Apparent motion of the field is then dominated by a translation due to the orbital motion of 4' s$^{-1}$, and field rotation can be ignored. The orientation of the spacecraft "nose" or spacecraft frame X-axis, relative to the commanded orbital path, is described by a roughly circular path ~1' in diameter with ~60 s period. The exact orientation at any given time is not well known, but the motion is smooth.

One of the simplest methods of image stabilization, "shift-and-add", records images over times short enough that no significant "smearing" of the image occurs in a single frame. Bright stars are later used to register the images so they may be co-added. A crude periodic "stepping" of the pointing, following the field motion, is required to keep the the field on the sensor during exposures. However, such strategies are only suitable for devices optimized for high frame-rate use, and not dominated by read noise or similar characteristics. This method is therefore not suitable for use with the NIR camera. The proposed HgCdTe detector is operated with Fowler sampling, which effectively samples the slope of a pixel many times during an exposure to reduce electronic noise by $N_{sample}^{1/2}$ (Beletic, 2008). If there is any significant change in the sky on the sensor pixels, i.e. "smearing", the noise is not reduced by this multiple sampling, and the performance is very poor. Therefore, true image stabilization is required for the science-critical NIR measurements.

### 6.2.ii  Feedback-controlled beam-steering stabilization

We required the mirror motors to be able to move the beam from the center nearly to the edge of the FOV of the X-ray camera, ~ 30 deg., in a few seconds. Therefore, moving 4' s$^{-1}$, the nominal motion of the zenith, is within the capability of the system. The challenge is in the precision control required for tracking, that is, stabilizing the pointing to follow the motion of the source relative to the instrument, with an error less than the size of our 2" pixels. It is now commonplace for commercial motors in a wide range of torques and masses to have 0.01" encoder resolution, so sufficiently precise knowledge of the motor shaft position is no problem. The required path of the beam (or motor shaft motion) relative to the instrument reference frame must also be known to great precision. This is more challenging: Typical spacecraft have uncertainties of the commanded vs. actual orientation of ~ 1 arc min, and orientation information is often not made available to the instruments in real time. However, the path of stars on fast-read sensors gives this information, and so can be used for feedback control of the mirror orientation.

The steering mirror is nominally moved to keep the image field centered on the image sensor for the expected orbital motion. Precise measurement of the image center drift relative to the expected orbital image motion will give the feedback required for high-resolution imaging. To achieve <1" tracking errors over 10 s exposures, drift of the field must be measured to << 1" over intervals spanning a small fraction of this time period. Requiring four measurements of 0.2" precision per 10 s would allow measurement of a tracking error of 0.28" in one interval, would allow application of four mirror shaft position/velocity profile corrections during the interval, and monitoring of tracking error and its derivative. Image PSF centroid measurements with error < 0.1 pixels are routinely made for SNR>8 sources. Measuring the uniform motion of stars on an image would allow us to average out the random part of this error, including that due to sub-pixel non-uniformities, so average field position errors of ≤ 0.05 pixels with > 4 stars is feasible. For our putative 0.08 sq. deg. field, we would require > 50 stars / sq. deg at high SNR. The SDSS catalog has ≥ 68 stars / sq. deg. for R≤14 at high latitudes. We estimate an EMCCD on a 30 cm telescope would detect a star R=14 @ 10 σ in < 20 ms, *greatly* exceeding these requirements, allowing more frequent and/or more precise tracking measurements. Coverage of a large field with optimal sensitivity may require a somewhat undersampled image scale for the science camera, and in this case, the final precision will be somewhat worse than our prediction. If the required precision cannot be achieved, this problem can be solved using a second EMCCD with an optimum pixel scale for centroid measurements on a smaller field.

The sequence of operational events after a GRB trigger would be as follows: When the X-ray camera detects a GRB, (a) the mirror is commanded to point the beam at the instantaneous position of the target relative to the X-ray camera coordinates, plus an offset to account for movement of the spacecraft orientation during the initial mirror pointing. The change in spacecraft orientation during the mirror pointing, due to the *planned* flight path is well known, but the actual orientation drift term is poorly known, changing by ~ $1"\ s^{-1}$. The error in the X-ray camera position contributes ~ 2-4' in the position prediction, and so the term from the uncertainty in spacecraft motion during the mirror move is insignificant. The beam is steered on target to within 2-4', with a 17' square FOV. (b) At this time the mirror motors are commanded to track at the approximate rate for the planned flight path. Fast EMCCD imaging commences, then (c) the feedback system is enabled, and PSF centroid motion provides control feedback. After a few feedback cycles, (d) the image is stabilized, and NIR imaging can commence, with << 1 pixel source centroid motion during 10 s or longer exposures, and sensitive NIR imaging results.

# TABLES

Table 1 *Swift* Performance Characteristics[1]

| | |
|---|---|
| [2]BAT Detections (yr$^{-1}$) | 94.5 |
| [3]BAT Real-Time Detections (yr$^{-1}$) | 86.6 |
| [4]UVOT Detections (yr$^{-1}$) | 30.4 |
| [5]UVOT Sensitivity in 10 s in W (5 σ; mag) | 18.75 |

(1) The above table gives values for selected periods of stable performance, noted in the footnotes below. More analysis of *Swift* measurement rates, without such selection, is given in Appendix Table A1 for comparison.

(2) 736 detections 2005 Mar. 18 through end of 2012 from *Swift* GRB Table (HEASARC).

(3) Same as above, eliminating Ground Analysis, MAXI, BATSS bursts (not removing image triggers).

(4) 234 detections 2005 Mar. 18 through end of 2012; from *Swift* GRB Table (HEASARC).

(5) From median reported 3 σ upper limit sensitivity in open or "W", filter, taken from a sample of GCN alerts between 8938 and 11019, scaled by $t^{1/2}$ to 10 s exposure, 5 σ. See Breeveld, Landsman, Holland, Roming, Kuin, & Page, 2011 for the calibration and definitions of UVOT filter bands.

**Table 2 BAT Data Trigger Analysis Selected Results**

| Area(cm$^2$) | N$_{detect}$(yr$^{-1}$) SNR≥5 | N$_{detect}$(yr$^{-1}$) SNR≥6.5 | N$_{detect}$(yr$^{-1}$) SNR≥8 |
|---|---|---|---|
| 191[a] | 42.7 | 32.5 | 27.5 |
| 800 | 70.3 | 61.2 | 52.2 |
| 1024[b] | 75.6 | 64.9 | 57.1 |
| 2511 | 81.3 | 78.9 | 74.4 |
| 5200 | 83.8 | 81.8 | 80.1 |

Results from simplified trigger and area scaling analysis of 224 bursts 2006 May 2 to 2008 October 7.

(a) Detecting area of UFFO UBAT (Kim, et al., 2012; but note that UFFO is in a much higher background orbit, which will adversely affect rates.)

(b) Approximate detecting area of proposed SVOM ECLAIRS (Godet, et al., 2012)

**Table 3** Swift GRB Sub-Samples and W Detection Rates

| | $\langle t_{mid} \rangle$ (s) | $t_{exp}$ (s) | $N_{BAT}^{(1)}$ Selected (Total) | $N_{UVOT}^{(2)}$ Early W Detections | UVOT Early W Detection Rate[3] (% [yr$^{-1}$]) |
|---|---|---|---|---|---|
| *060502a-081007 | $t_{mid} < 170s$ | 144 | ~98 | 207 (224) | 44 | 21.2 [18.1] |
| 081008-121229a | $t_{mid} < 220s$ | 179 | ~147 | 357 (379) | 67 | 18.8 [15.8] |

(*) The UVOT Early Brightness Distribution Sample

(1) $N_{BAT}$ Selected includes only bursts detected by our simple algorithm, and only real-time rate triggers, as other types of triggers do not permit rapid optical follow-up by UVOT. (The selection logic is given by: (not an image trigger) and (not a ground processing detection) and (not a failure of our trigger algorithm).)

(2) $N_{UVOT}$ gives the number of detections for the early sample selection ($t_{mid} \leq 170$ s and BAT selected; see following note).

(3) For rate calculation, the time period for the first sample is 2006 May 2.0 to 2008 Oct. 8. As there were bursts on the days before and after, and the number of days in the sample time is many hundred, we did not strive for greater than 1 day accuracy. For the second time period, the begin time is 2008 Oct. 8.0, given a burst the day before; the end time was the "blind" pre-selected, date 2013 Jan 1.0.

**Table 4** Annual Detection Rate Early Sample Lower Limits As A Function of X-ray and Optical Instrument Characteristics for 6.5 Sigma Trigger

| | | X-ray Detecting Area (cm$^2$) | | | | |
|---|---|---|---|---|---|---|
| | | 191$^{UF}$ | 800 | 1024$^{SE}$ | 2511 | 5200$^{SB}$ |
| Optical Aperture Diameter (cm) / detector | 10/ICCD | 5.2 | 6.6 | 6.6 | 7.0 | 7.0 |
| | 10/CCD | 8.9 | 11.2 | 11.2 | 11.9 | 12.0 |
| | 30/ICCD | 7.7 | 11.8 | 11.8 | 12.7 | 12.7 |
| | 30/CCD | 9.4 | 14.4 | 14.4 | 15.4 | 15.5 |

UF= Detecting area of UFFO UBAT (Kim, et al., 2012; but note that UFFO is in a much higher background orbit.)

SE = Approximate detecting area of proposed SVOM ECLAIRS (Godet, et al., 2012)

SB = Detecting area of *Swift* BAT (Gehrels, et al., 2004)

ICCD = UVOT photocathode microchannel plate intensified CCD

**Table 5 Key Technologies**

| Detector | Principle | λ or E | QE (%) | Comment (Example) |
|---|---|---|---|---|
| microchannel plate intensified CCD | photoelectric effect in metallic photocathode | ~0.1-0.6μm | <25 | fast reads approximate photon-counting device (UVOT) |
| CCD (charged-coupled device) | collection of photoelectrons in silicon | 0.41-0.90 μm | >90 | Current standard integrating device (ACS, STIS, ground-based optical cameras.) |
| EMCCD (electron multiplying CCD) | same, with electrons multiplied *before* read, so read noise negligible | "" | "" | Fast reads approximate photon-counting device with CCD performance (CHIMERA) |
| H2RG NIR array | HgCdTe-doped Hybrid CMOS semiconductor allows collection of photoelectrons | 0.8-2.5 μm (up to 5μm possible) | >80 | integrating device<br><br>short exposure noise pentalty<br><br>sensitive to thermal background of optical system (NICMOS) |
| CZT | Collection of photoelectrons in Cd-Zn-Te semiconductor | 15-150 keV | >38% (15-100 keV) | Current standard (*Swift*/BAT, Integral/IBIS) |
| Cooled, low-energy sensitive CdTe or CZT | Higher-resistivity crystals than above, cooled to ~−20C (1) | 5- 150 keV | Same, extending to lower energy | Low energy response improved through cooling, crystal growth, and improved readout electronics. (SVOM/ECLAIRS) |

**Other Technologies**

| Technology | Purpose | Space Heritage or Example | Comment |
|---|---|---|---|
| Beam-steering mirror | Redirect telescope beam in ~ 1s | Scaning mirrors, e.g., Aqua, GOES imager, Spitzer MIPS, commercial gimbal mirrors | Space astronomical gimbal mirror demonstrated in lab (2) |
| Mirror cooling | Reduce thermal background | WISE, Spitzer | Cryogen requirements decrease as passive cooling design improves (e.g. changes from *SIRTF* to Spitzer design) |

[1] (Krawczynski, Jung, Perkins, Burger, & Groza, 2004); Laurent, Philippe (CEA), 2012, private communication.
[2] (Jeong, et al., 2013)

**Table 6 Annual Detection Rate Estimates (Lower Limits) As A Function of X-ray and NIR Instrument Characteristics**

| | | X-ray Detecting Area (cm$^2$) | | | | |
|---|---|---|---|---|---|---|
| | | 191$^{UF}$ | 800 | 1024$^{ES}$ | 2511 | 5200$^{SB}$ |
| NIR Exposure Time/Aperture | 10 s exposure 10 cm – 30 cm Aperture | 12 | 19 | 19 | 20 | 20 |
| | 100 s exposure 30 cm Aperture | 13 | 20 | 20 | 22 | 22 |

UF= detector area of UFFO UBAT (Kim, et al., 2012)

ES = approximate detector area of SVOM ECLAIRS (Godet, et al., 2012)

SB = the detector area of *Swift* BAT (Gehrels, et al., 2004)

The values in the table above give lower limits to annual detection rates for the given exposures. These results are scaled from the results in Table 4, i.e. we approximated extinction as being uncorrelated with BAT properties, approximating a fixed fractional increase from the number of unextinguished bursts for each X-ray detection area.